\documentclass[11pt, a4paper]{article}

\usepackage[utf8]{inputenc}
\usepackage[T1]{fontenc}
\usepackage{lmodern}      
\usepackage{microtype}    
\usepackage{amsmath, amssymb, amsthm}
\usepackage{graphicx}
\usepackage{booktabs}
\usepackage[margin=1in]{geometry}
\usepackage{hyperref}
\usepackage{xcolor}
\usepackage{listings}
\usepackage{caption}
\usepackage{subcaption}

\hypersetup{
    colorlinks=true,
    linkcolor=blue,
    filecolor=magenta,      
    urlcolor=cyan,
    citecolor=red,
    breaklinks=true
}

\definecolor{codegreen}{rgb}{0,0.6,0}
\definecolor{codegray}{rgb}{0.5,0.5,0.5}
\definecolor{codepurple}{rgb}{0.58,0,0.82}
\definecolor{backcolour}{rgb}{0.95,0.95,0.92}

\lstdefinestyle{mystyle}{
    backgroundcolor=\color{backcolour},   
    commentstyle=\color{codegreen},
    keywordstyle=\color{magenta},
    numberstyle=\tiny\color{codegray},
    stringstyle=\color{codepurple},
    basicstyle=\ttfamily\footnotesize,
    breakatwhitespace=false,         
    breaklines=true,                 
    captionpos=b,                    
    keepspaces=true,                 
    numbers=left,                    
    numbersep=5pt,                  
    showspaces=false,                
    showstringspaces=false,
    showtabs=false,                  
    tabsize=2
}
\title{\textbf{Dynamics of Socio-Institutional Asynchrony in Generative AI: Analyzing the Relative Importance of Intervention Timing vs. Enforcement Efficiency via the Socio-Institutional Asynchrony Model (SIAM)}}
\author{Taeyoon Kim \\ \textit{Independent Researcher} \\ \texttt{kimyoontae0213@gmail.com}}
\date{\today}

\begin{document}

\maketitle

\begin{abstract}
The \textbf{super-exponential} growth of Generative AI has exacerbated the ``Institutional Mismatch'' between the speed of technological diffusion and the pace of institutional adaptation. This study proposes the \textbf{Socio-Institutional Asynchrony Model (SIAM)} to quantitatively evaluate the effectiveness of two policy levers: \textbf{Intervention Timing} versus \textbf{Enforcement Efficiency}. Based on the timeline of the EU AI Act and the compute doubling time of 6 months, we conducted a high-precision simulation ($N=10,001$). The results indicate that the ``Timing Strategy'' (starting earlier) reduces the social burden by approximately \textbf{64.2\%}, whereas the ``Efficiency Strategy'' (shortening the rollout period) reduces it by only \textbf{30.3\%}. We \textbf{analytically demonstrate} that advancing the start timing has a \textbf{structurally higher sensitivity} (approx. 2.1x relative effectiveness) in reducing the regulatory vacuum compared to increasing enforcement speed. These findings suggest that the core value of AI governance lies in \textbf{Proactive Timeliness} rather than reactive administrative efficiency.
\end{abstract}

\section{Introduction}

\subsection{Background: The 5th Technological Revolution and Crisis}
According to techno-economic scholar Carlota Perez, every major technological transition undergoes a ``Turning Point'' crisis—a turbulent period between the initial \textit{Installation Period} of creative destruction and the subsequent \textit{Deployment Period} of widespread benefit \cite{perez2002}. 

This phenomenon fundamentally creates a structural gap, a concept deeply rooted in Ogburn's Cultural Lag theory (1922), which posits that material culture (technology) advances faster than non-material culture (institutions), causing inevitable social maladjustment \cite{ogburn1922}. Currently, Generative AI is exhibiting explosive growth, with training compute doubling approximately every 6 months \cite{sevilla2022}, reaching the peak of its Installation Period. However, institutional frameworks remain stagnant in past paradigms, exacerbating this structural mismatch.

\subsection{Objective}
This study translates Perez's qualitative theory and Ogburn's lag concept into a mathematical model (SIAM) to compare the relative effectiveness of two policy levers:
\begin{itemize}
    \item \textbf{Timing Strategy ($L$):} Advancing the start of legislative discussions.
    \item \textbf{Efficiency Strategy ($D$):} Accelerating the enforcement and rollout of established regulations.
\end{itemize}

\vspace{0.3cm}
\noindent \textbf{Scope of Study:} It is crucial to note that this study aims to elucidate the \textit{structural trends} and \textit{dynamic mechanisms} of socio-institutional asynchrony, rather than providing precise empirical predictions of specific future events. By utilizing the SIAM framework, we focus on identifying the relative sensitivity of key variables—specifically, comparing the structural impact of intervention timing versus enforcement efficiency on the total social burden.

\section{Theoretical Framework}

We adopt the \textbf{Logistic Growth Model} as a mathematical necessity to represent the life-cycle mechanism of the Techno-Economic Paradigm proposed by Perez.

\subsection{Mapping Perez's Phases to Logistic Function}
The diffusion of technology and institutions generally follows a Sigmoid path, a phenomenon widely characterized by \textbf{Rogers' Diffusion of Innovations theory \cite{rogers2003}}. To model this phenomenological pattern mathematically, we utilize the Verhulst equation:
\begin{equation}
    \frac{dy}{dt} = k \cdot y \cdot (1 - y)
\end{equation}

\begin{itemize}
    \item \textbf{Installation Period (Positive Feedback, $y'' > 0$):} Matches the convex region where adoption accelerates ($y \ll 1$). While Rogers describes this as the early adopter phase, in the context of Generative AI, this phase is driven by rapid compute growth \cite{sevilla2022}, acting as a potent exogenous force.
    \item \textbf{Turning Point (Inflection Point, $y'' = 0$):} Corresponds to $t_0$ where growth speed peaks. This is the critical moment of maximum tension between the surging technology and lagging institutions.
    \item \textbf{Deployment Period (Negative Feedback, $y'' < 0$):} Matches the concave region where the system stabilizes.
\end{itemize}

\section{Mathematical Modeling}

All time variables are in \textbf{Years}, and diffusion levels are \textbf{Dimensionless (0--1)}. Variables are summarized in Table \ref{tab:variables}.

\subsection{Variable Definitions}
\begin{table}[h]
\centering
\begin{tabular}{clcl}
\toprule
\textbf{Symbol} & \textbf{Definition} & \textbf{Unit} & \textbf{Note} \\
\midrule
$S(t)$ & Tech Diffusion Level & - & $S \in [0, 1]$ \\
$I(t)$ & Institutional Coverage & - & $I \in [0, 1]$ \\
$L$ & Legislation Lag (Start) & Year & Assumed start of rollout ($I \approx 0.1$) \\
$D$ & Rollout Duration & Year & Duration from 10\% to 90\% coverage \\
$n$ & Institutional Speed & Year$^{-1}$ & Slope constant ($n \propto 1/D$) \\
$t_{0,I}$ & Institutional Inflection & Year & Midpoint where $I(t_{0,I}) = 0.5$ \\
\bottomrule
\end{tabular}
\caption{Variable Definitions}
\label{tab:variables}
\end{table}

\subsection{Technological Pressure and Societal Permeability}
\textbf{Step 1: Raw Growth Rate $k_S$} \\
Using the doubling time $T_d = 0.5$ years \cite{sevilla2022}:
\begin{equation}
    e^{k_S \cdot T_d} = 2 \implies k_S = \frac{\ln 2}{T_d} \approx 1.386
\end{equation}

\textbf{Step 2: Societal Permeability Coefficient ($\alpha$)} \\
We define the effective societal pressure, $S_{calibrated}(t)$, as the product of the raw infrastructure growth $S(t)$ and a \textit{Societal Permeability Coefficient} $\alpha$:
\begin{equation}
    S_{calibrated}(t) = \alpha \cdot \frac{1}{1 + e^{-k_S(t - 1.0)}}
\end{equation}
where $S(t)$ represents the raw growth of technological capability (e.g., computational power), and $\alpha$ ($0 < \alpha \le 1$) represents the efficiency with which this capability translates into societal impact.

\textbf{Assumption:} While $\alpha$ typically assumes a value less than 1.0 due to adoption friction, in this study, considering the ``Frenzy'' phase of the current AI boom where infrastructure growth directly drives market expectations \cite{perez2002}, we assume $\mathbf{\alpha \approx 1.0}$. This allows us to simulate the \textit{maximum pressure scenario}. Thus, hereafter $S(t)$ refers to $S_{calibrated}(t)$.

\subsection{Derivation of Institution Function I(t)}
The modeling of institutional response $I(t)$ as a sigmoid curve is grounded in Rogers' Diffusion of Innovations theory \cite{rogers2003}. Just as technology adoption follows a bell-shaped distribution of social acceptance—accumulating into an S-curve—institutional formation also represents a diffusion process of \textit{social consensus} among policymakers and the public. Therefore, it is structurally valid to model the institutional maturity level as a logistic growth function.

\vspace{0.3cm}
\noindent \textbf{Step 1: Relation between Speed $n$ and Duration $D$} \\
Defining $D$ as the time to go from $y=0.1$ to $y=0.9$, and using the Logit transformation:
\begin{equation}
    n \cdot D = \text{logit}(0.9) - \text{logit}(0.1) = \ln(9) - \ln(1/9) = 2\ln 9
\end{equation}
\begin{equation}
    \therefore n = \frac{2\ln 9}{D}
\end{equation}

\textbf{Step 2: Dynamic Definition of Inflection Point $t_{0,I}$} \\
Assuming the start lag $L$ maps to the 10\% diffusion point, the midpoint is shifted by half the duration:
\begin{equation}
    t_{0,I} = L + \frac{D}{2}
\end{equation}

\textbf{Step 3: Final Function}
\begin{equation}
    I(t) = \frac{1}{1 + e^{-\frac{2\ln 9}{D}\left(t - \left(L + \frac{D}{2}\right)\right)}}
\end{equation}

\subsection{Structural Sensitivity Analysis}
We analyze the partial derivatives of the inflection point $t_{0,I}$ with respect to $L$ and $D$.

\begin{align}
    \text{Sensitivity to Timing ($L$):} & \quad \frac{\partial t_{0,I}}{\partial L} = 1 \\
    \text{Sensitivity to Efficiency ($D$):} & \quad \frac{\partial t_{0,I}}{\partial D} = 0.5
\end{align}

\textbf{Interpretation:}
\begin{itemize}
    \item Reducing the start lag ($L$) shifts the inflection point by the full amount (100\%).
    \item Reducing the duration ($D$) only shifts it by half (50\%).
    \item Thus, the \textbf{Timing Strategy has structurally higher sensitivity} ($2\times$) compared to the Efficiency Strategy.
\end{itemize}

\subsection{Calculation of Total Social Burden}
The total social burden ($H_{total}$) is the cumulative time-gap integral defined as:
\begin{equation}
    H_{total} = \int_{0}^{T_{max}} \max(S(t) - I(t), 0) \, dt
\end{equation}

\textbf{Dimension Analysis:} The unit of $H_{total}$ is $[\text{Year} \cdot \text{Normalized Gap}]$, representing the cumulative duration of the regulatory vacuum weighted by the intensity of the mismatch.

\section{Experimental Setup}

\subsection{Calibration (EU AI Act Case)}
\begin{itemize}
    \item \textbf{Legislation Lag ($L \approx 1.67$ yr):} Calculated from ChatGPT launch (Nov 30, 2022) \cite{openai_chatgpt} to EU AI Act Entry into Force (Aug 01, 2024) \cite{eu_ai_act}. Duration is approx. 610 days $\approx 1.67$ years.
    \item \textbf{Rollout Duration ($D \approx 2.0$ yr):} Based on Article 113 of the EU AI Act, General Application applies 24 months (2 years) after entry into force \cite{eu_ai_act}.
    \item \textbf{Baseline Inflection ($t_{0,I}$):} $1.67 + 1.0 = 2.67$ years.
\end{itemize}

\subsection{Methodology}
\begin{itemize}
    \item \textbf{Time Domain:} $t \in [0, 15]$ years.
    \item \textbf{Resolution:} $N=10,001$ grid points (Odd number for Simpson's rule).
    \item \textbf{Algorithm:} \texttt{scipy.integrate.simpson} (guaranteeing $O(\Delta t^4)$ accuracy \cite{burden2010}).
\end{itemize}

\section{Simulation Results}

\subsection{Scenario Comparison}
As shown in Table \ref{tab:scenarios}, we compared three distinct strategies.

\begin{table}[h]
\centering
\begin{tabular}{lcccc}
\toprule
\textbf{Scenario} & \textbf{Parameters} & \textbf{$\Delta t_{0,I}$} & \textbf{Burden Reduction} \\
\midrule
(A) Baseline & $L=1.67, D=2.0$ & - & Reference \\
(B) Timing & $\mathbf{L=0.67}, D=2.0$ & \textbf{-1.0 yr} & \textbf{64.2\%} \\
(C) Efficiency & $L=1.67, \mathbf{D=1.0}$ & \textbf{-0.5 yr} & \textbf{30.3\%} \\
\bottomrule
\end{tabular}
\caption{Simulation Scenarios and Results}
\label{tab:scenarios}
\end{table}

\begin{figure}[h]
    \centering
    \includegraphics[width=\textwidth]{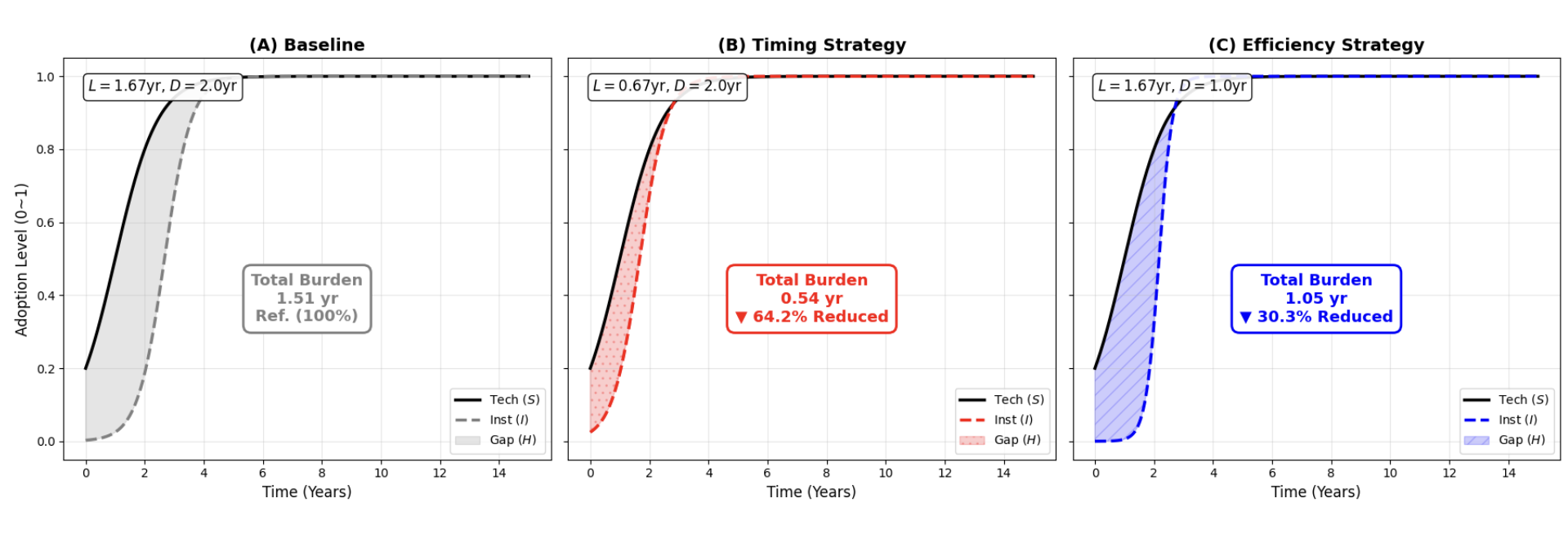}
   \caption{Comparison of Total Social Burden ($H_{total}$) between Timing and Efficiency Strategies. Note that the Y-axis represents the \textbf{Maturity Level of Adoption/Institution (0--1)}. The red shaded area corresponds to the Timing Strategy, and the blue shaded area corresponds to the Efficiency Strategy.}
    \label{fig:results}
\end{figure}

\subsection{Analysis}
\begin{enumerate}
    \item \textbf{Efficiency Strategy:} Even with a 2x increase in administrative speed ($\Delta D = -1.0$), the burden reduction was limited to approx. \textbf{30.3\%}.
    \item \textbf{Timing Strategy:} Starting 1 year earlier ($\Delta L = -1.0$) resulted in a substantial burden reduction of approx. \textbf{64.2\%}.
    \item \textbf{Conclusion:} The Timing Strategy demonstrated \textbf{higher effectiveness} (approx. 2.1x ratio) compared to the Efficiency Strategy. This confirms the theoretical sensitivity analysis, where the leverage of timing is structurally superior.
\end{enumerate}

\section{Conclusion}

This study quantitatively explored the dynamics of the ``Turning Point'' crisis in the era of Generative AI through the Socio-Institutional Asynchrony Model (SIAM). By integrating the theoretical frameworks of Perez and Ogburn with a mathematical sensitivity analysis, we examined the relative efficacy of intervention timing versus enforcement efficiency.

Our simulation indicates that the \textbf{Timing Strategy} yields a social burden reduction of approximately \textbf{64.2\%}, notably higher than the \textbf{30.3\%} reduction achieved by the \textbf{Efficiency Strategy}. However, it is crucial to note that since this model is calibrated on theoretical assumptions rather than extensive historical empirical datasets, these specific figures should be interpreted as demonstrative of underlying structural trends rather than as precise predictive metrics. They primarily illustrate the inherent mechanics where the institutional inflection point is structurally twice as sensitive to the start timing ($L$) as it is to the rollout duration ($D$).

These findings offer significant implications for addressing the \textbf{Collingridge Dilemma} in AI governance. While administrative efficiency is undoubtedly necessary, our model suggests that it is insufficient to counterbalance the super-exponential diffusion of AI if the initial intervention is delayed. Consequently, policymakers should consider a strategic pivot: rather than incurring significant delays to craft a flawless legal framework, establishing a preliminary regulatory ``zero-point''—potentially through soft law or guidelines—may be a more pragmatic approach to mitigating the cumulative social costs of the institutional void.

Furthermore, considering that the EU represents a 'fast-adopter' case in the global regulatory landscape, the simulated burden in this study likely represents a lower bound for the global average. For nations like South Korea or the US, where the legislative clock is still ticking ($L > 1.67$), the 'Cost of Inaction' is accumulating at a structurally faster rate. Therefore, policymakers in these jurisdictions should prioritize minimizing $L$ through immediate soft-law interventions rather than waiting for a perfect legislative consensus.

Finally, this study serves as a foundational framework that bridges theoretical modeling with potential data-driven applications. Future research can significantly advance this model by incorporating empirical datasets from historical technological disruptions—such as the adoption curves of the internet or mobile communications—to calibrate the diffusion parameters ($k_S$) and institutional lags ($L, D$) with greater precision. Such empirical validation will not only test the robustness of the structural sensitivities identified in this study but also enable a more granular and detailed analysis of specific regulatory scenarios. Ultimately, evolving SIAM with real-world data will transform it into a comprehensive empirical tool for evidence-based policymaking.

\newpage
\newpage

\bibliographystyle{plain}

\begin{thebibliography}{99}

\bibitem{perez2002}
Perez, C. (2002). 
\textit{Technological Revolutions and Financial Capital: The Dynamics of Bubbles and Golden Ages}. 
Edward Elgar Publishing.

\bibitem{ogburn1922}
Ogburn, W. F. (1922).
\textit{Social Change with Respect to Culture and Original Nature}.
B. W. Huebsch.

\bibitem{rogers2003}
Rogers, E. M. (2003). 
\textit{Diffusion of Innovations} (5th ed.). 
Free Press.

\bibitem{sevilla2022}
Sevilla, J., et al. (2022). 
``Compute Trends Across Three Eras of Machine Learning''. 
\textit{arXiv preprint arXiv:2202.05924}.

\bibitem{burden2010}
Burden, R. L., \& Faires, J. D. (2010). 
\textit{Numerical Analysis} (9th ed.). 
Brooks/Cole.

\bibitem{openai_chatgpt}
OpenAI. (2022). 
``Introducing ChatGPT''. 
\textit{OpenAI Official Blog}. Available: \url{https://openai.com/blog/chatgpt}. [Accessed: Dec. 20, 2025].

\bibitem{openai_devday}
OpenAI. (2023). 
``New models and developer products announced at DevDay''. 
\textit{OpenAI Official Blog}. Available: \url{https://openai.com/blog/new-models-and-developer-products-announced-at-devday}. [Accessed: Dec. 20, 2025].

\bibitem{eu_ai_act}
European Union. (2024). 
``Regulation (EU) 2024/1689 of the European Parliament and of the Council''. 
\textit{Official Journal of the European Union}, L, 2024/1689. Available: \url{https://eur-lex.europa.eu/eli/reg/2024/1689/oj}.

\end{thebibliography}

\newpage
\appendix
\section{Simulation Code (Python)}
\label{appendix:code}

The following Python code generates the simulation results and Figure \ref{fig:results}.

\begin{lstlisting}[language=Python]
import numpy as np
import matplotlib.pyplot as plt
from scipy.integrate import simpson

def run_siam_visual_simulation():
    # [1. Settings] High precision setup
    N_GRID = 10001 
    T_MAX = 15
    t = np.linspace(0, T_MAX, N_GRID)
    
    # [2. Tech Model S(t)]
    # Sevilla et al. (2022): Doubling time ~ 0.5 yr
    T_d = 0.5  
    S_k = np.log(2) / T_d
    S_t0 = 1.0 
    S_curve = 1 / (1 + np.exp(-S_k * (t - S_t0)))
    
    # [3. Institution Model I(t)]
    def get_institution_curve(L, D):
        # Slope n derived from Logit(0.9) - Logit(0.1)
        n = (2 * np.log(9)) / D
        # Midpoint t0 derived from L + D/2
        t0_I = L + (D / 2)
        return 1 / (1 + np.exp(-n * (t - t0_I)))

    # [4. Scenarios]
    scenarios = [
        {"id": "(A) Baseline", "L": 1.67, "D": 2.0, "color": "gray", "hatch": ""},
        {"id": "(B) Timing Strategy", "L": 0.67, "D": 2.0, "color": "red", "hatch": ".."},
        {"id": "(C) Efficiency Strategy", "L": 1.67, "D": 1.0, "color": "blue", "hatch": "//"}
    ]
    
    # [5. Simulation & Visualization]
    fig, axes = plt.subplots(1, 3, figsize=(18, 6), sharey=True)
    
    # Calculate Baseline Area
    I_base = get_institution_curve(1.67, 2.0)
    gap_base = np.maximum(S_curve - I_base, 0)
    base_area = simpson(y=gap_base, x=t)

    print(f"--- Simulation Results ---")
    print(f"Baseline Area: {base_area:.4f}")

    for i, sc in enumerate(scenarios):
        ax = axes[i]
        
        # Generate Curve
        I_curve = get_institution_curve(sc["L"], sc["D"])
        
        # Calculate Burden (Area)
        gap = np.maximum(S_curve - I_curve, 0)
        area = simpson(y=gap, x=t)
        
        # Calculate Reduction
        reduction = (base_area - area) / base_area * 100
        impact_text = "Ref. (100%)" if i == 0 else f"-{reduction:.1f}% Reduced"
        
        # Print Text Result
        if i > 0:
            print(f"{sc['id']}: Area={area:.4f}, Reduction={reduction:.2f}%")

        # Plotting
        ax.plot(t, S_curve, 'k-', lw=2.5, label='Tech ($S$)')
        ax.plot(t, I_curve, color=sc['color'], ls='--', lw=2.5, label='Inst ($I$)')
        ax.fill_between(t, S_curve, I_curve, where=(S_curve > I_curve),
                        color=sc['color'], alpha=0.2, hatch=sc['hatch'],
                        label='Gap ($H$)')
        
        # Styling
        ax.set_title(sc['id'], fontsize=14, fontweight='bold')
        param_str = f"$L={sc['L']}$yr, $D={sc['D']}$yr"
        ax.text(0.05, 0.95, param_str, transform=ax.transAxes, 
                fontsize=12, va='top', bbox=dict(boxstyle="round", fc="white", alpha=0.9))
        
        # Result Box
        bbox_props = dict(boxstyle="round,pad=0.5", fc="white", ec=sc['color'], lw=2)
        ax.text(0.5, 0.4, f"Total Burden\n{area:.2f} yr\n{impact_text}", 
                transform=ax.transAxes, ha='center', va='center', 
                fontsize=13, fontweight='bold', color=sc['color'], bbox=bbox_props)

        ax.grid(True, alpha=0.3)
        ax.set_xlabel("Time (Years)")
        if i == 0: ax.set_ylabel("Adoption Level (0~1)")
        ax.legend(loc='lower right')

    plt.suptitle("Comparative Analysis: Timing ($L$) vs. Efficiency ($D$) Strategies", fontsize=16, y=1.02)
    plt.tight_layout()
    plt.savefig('siam_results.png', dpi=300, bbox_inches='tight') # Save for LaTeX
    plt.show()

if __name__ == "__main__":
    run_siam_visual_simulation()
\end{lstlisting}

\end{document}